\documentclass[useAMS,usenatbib]{mn2e}
\usepackage{txfonts}
\usepackage{natbib}
\usepackage{aas_macros}
\usepackage{graphicx}
\usepackage{subfigure}
\usepackage{mathrsfs}
\usepackage{xfrac}

\newcommand{\gsim}{\mathrel{\hbox{\rlap{\lower.55ex \hbox {$\sim$}}
                   \kern-.3em \raise.4ex \hbox{$>$}}}}
\newcommand{\lsim}{\mathrel{\hbox{\rlap{\lower.55ex \hbox {$\sim$}}
                   \kern-.3em \raise.4ex \hbox{$<$}}}}

\newcommand\sdensity{$\rm g \: cm^{-2}$}

\newcommand\rorbit{$r_{\rm p}$}

\newcommand\solarmass{$\rm M_{\odot}$}

\newcommand\earthmass{$\rm M_{\oplus}$}

\title[Planetesimal accumulation near gaps]{On the accumulation of planetesimals near disc gaps created by protoplanets}
\author[B.A. Ayliffe et al.]
{Ben A. Ayliffe$^{1,2}$\thanks{E-mail:ayliffe@astro.ex.ac.uk},
Guillaume Laibe$^2$,
Daniel J. Price$^2$ and
Matthew R. Bate$^{1,2}$\thanks{E-mail:mbate@astro.ex.ac.uk}\\
$^1$School of Physics, University of Exeter, Stocker Road, Exeter EX4 4QL\\
$^2$Monash Centre for Astrophysics (MoCA) \& School of Mathematical Sciences, Monash University, Clayton, Vic 3800, Australia}

\date{\today}
\begin{document}
\maketitle

\begin{abstract}
We have performed three-dimensional two-fluid (gas-dust) hydrodynamical models of circumstellar discs with embedded protoplanets ($3 - 333$~\earthmass) and small solid bodies (radii 10cm to 10m). We find that high mass planets ($\gsim$ Saturn mass) open sufficiently deep gaps in the gas disc such that the density maximum at the outer edge of the gap can very efficiently trap metre-sized solid bodies. This allows the accumulation of solids at the outer edge of the gap as solids from large radii spiral inwards to the trapping region. This process of accumulation occurs fastest for those bodies that spiral inwards most rapidly, typically metre-sized boulders, whilst smaller and larger objects will not migrate sufficiently rapidly in the discs lifetime to benefit from the process. Around a Jupiter mass planet we find that bound clumps of solid material, as large as several Earth masses, may form, potentially collapsing under self-gravity to form planets or planetesimals. These results are in agreement with \cite{LyrJohKlaPis2009}, supporting their finding that the formation of a second generation of planetesimals or of terrestrial mass planets may be triggered by the presence of a high mass planet.

\end{abstract}


\section{Introduction}

Despite its diminutive scale, the metre-barrier in planetesimal growth remains a significant hurdle in the development of a cohesive theory of planet formation. In the early seventies it was realised that solids suspended in a circumstellar disc would settle towards the midplane, and it was suggested that the resulting density increase could bring about an instability within the layer \citep{Lyt1972, PolFri1972, GolWar1973}. \cite{GolWar1973} calculated that such a gravitational instability could lead to planetesimal growth from sub-cm scales to 100m scales in one generation, and occur on a timescale of just a few years. Subsequent generations could then grow to multi-kilometre scales, forming the progenitors necessary for terrestrial planets and the cores of giant planets. All that was required for this process to work was a high density being achieved in the dust layer ($\rho_{\rm  d,critical}$) through settling, which corresponds to a thin layer ($H_{\rm d} \approx \Sigma_{\rm d}/\rho_{\rm d,critical}$) born of low velocity dispersions ($\sigma_{\rm d}$) within the dust population ($H_{\rm d} \approx \sigma_{\rm d}/\Omega_{\rm d}$); $H_{\rm d}$ is the scaleheight of the dust layer, $\Sigma_{\rm d}$ is the surface density of this layer, and $\Omega_{\rm d}$ is the Keplerian angular velocity.

Credence was lent to the \cite{GolWar1973} model in light of \citeauthor{Wei1977}'s calculations regarding aerodynamic drag on planetesimals. These showed extremely rapid infall for metre-scale bodies (lifetime $< 100$ orbits from 1~au, \citealt{Wei1977}) such that skipping over the $\mathcal{O}({\rm 1m})$ scale in planetesimal growth seemed necessary. However, \citeauthor{Wei1980} went onto show that a layer of dust which became sufficiently dense ($\Sigma_{\rm d} \gtrsim \Sigma_{\rm g}$) would exert a significant drag on the adjoining gas layers. This drag would accelerate the sub-Keplerian velocities of the pressure supported gas up to the Keplerian values of the dust layer. The resulting vertical stratification of the gas velocity above and below the dust disc would lead to turbulent shear layers, where the gas is stirred to velocities of several metres per second (\citealt{Wei1980}, \citealt*{CuzDobCha1993}). In turn, the coupling through drag of the dust to the gas leads to these velocity dispersions being established in the dust layer, and so a self-induced limit to the scaleheight of the dust layer exists. \cite{CuzDobCha1993} showed that this limit is such that solid bodies up to 1 metre in size are unlikely to decouple sufficiently from the gas to enable a gravitational collapse to occur. Additionally, the rapid radial infall of metre-scale planetesimals due to drag was found to generate sufficient velocity dispersions in such a population of bodies as to delay the onset of any instability until they reached scales of $10-100$~m \citep{Wei1995}. However, growth through agglomeration to these scales is challenging due to both difficulties in the sticking efficiency of metre-body collisions \citep{BluWur2008}, and the depletion of these objects through drag-induced infall into their stars, depleting the available material \citep{Wei1977}.

An initially quiescent disc in which grains underwent vertical settling had been established as an unlikely environment in which to build planetesimals through gravitational instability. So from the mid-nineties onwards research began to focus on what conditions might allow higher densities to be achieved, densities that could trigger a gravitational collapse or could accelerate collisional growth to a practicable extent. One suggestion, based on the lapsed theories of \cite{Wei1943}, is that dust might become trapped in and amongst long lived vortices in the circumstellar disc, leading to just such high density concentrations (\citealt{BarSom1995, TanBabDubPro1996, HodBra1998, Cha2000, GodLiv2000, SupLin2000, de-Bar2001}; \citealt*{JohAndBra2004}; \citealt{BarMar2005, HenKen2010}). As well as vortices in the plane of the disc, \cite{KlaHen1997} found that millimetre scale particles can be effectively trapped and concentrated within eddies with rotation axes parallel to the disc midplane, e.g. convection cells.

It is also possible that circumstellar discs with metallicity much higher than solar composition could `saturate' the Kelvin-Helmholtz unstable layer, essentially providing more solid material than can effectively be stirred up by the turbulent flow \citep{YouShu2002}. Those particles not stirred up could settle to form a high-density layer prone to gravitational instability. This high ratio of solids surface density relative to the gas could also be achieved through accumulation in the dust layer, for example by differential radial drift \citep{YouChi2004, GarLin2004}. However, \cite{GomOst2005} showed that the inclusion of Coriolis forces leads to the onset of self-induced turbulence for thicker discs (higher Richardson numbers than assumed in the works just discussed), requiring much higher solids surface density augmentation to trigger instabilities.

\cite{Whi1972} found that dust and planetesimals are driven towards pressure maxima within a gas disc. This provides an additional route by which to build high concentrations of solid material and trigger growth to larger scales. \cite{RicLodPriArm2004,RicLodPriArm2006} demonstrated using numerical models that concentrations forming in spiral arms could collapse under self-gravity, building larger gravitationally bound objects. A further consolidation of various ideas was presented by \cite{JohOisMacKla2007} who found that the migration of metre-size bodies to pressure maxima in the gas disc could lead to a runaway process of growth, dubbed the streaming instability. Unlike the quiescent process of solids settling to the midplane, the streaming instability benefits from drag exerted upon the gas by the large mass of solid bodies. Accelerating the local gas to near Keplerian gas velocities reduces the headwind felt by the solids, so reducing their inward drift and allowing solids drifting inwards from larger radii to collect in the overdense region. This process may lead to favourably high solids surface densities, sufficient to trigger gravitational collapse (\citealt*{JohHenKla2006}; \citealt{JohOisMacKla2007,LyrJohKlaPis2009}), as found once self-gravity was included \citep{JohYouMac2009}.

Other studies have considered the evolution of dust in a circumstellar disc in the presence of a protoplanet. \cite{PaaMel2004, PaaMel2006a}, \cite*{MadFouGon2007} and \cite{FouMadGonMur2007} found that gap formation about a planet was more rapid in the dust population than in the gas. They also found that gap formation in the dust layer was able to occur for much lower planet masses than gap formation in the gas. \cite{FouMadGonMur2007} also point out that the accumulation of dust at the edge of a planetary gap might lead to higher densities, and aid in the formation of subsequent planets. \cite{Paa2007} investigated the accretion of dust and boulders onto a growing protoplanet using a two-fluid gas-solids hydrodynamics model. He found that boulders migrating inwards through a smooth gas disc due to the headwind could become trapped in resonance with the protoplanet, halting their inward migration and leading to high densities of solid material. In an evolving gas disc, where the protoplanet has opened a gap, it was observed that the pressure maxima at the gap edges could effectively halt the inward drift of dust, preventing it from reaching the growing planet, and allowing considerable accumulations to form. Subsequently \cite{LyrJohKlaPis2009} examined whether such sites could lead to the growth of planetesimals and planetary embryos through gravitational collapse. Their models led to the formation of a whole range of bodies, ranging from lunar mass objects to multi-Earth mass planetary embryos. It therefore appears possible that an extant giant planet can play an important role in triggering further planetesimal and planet growth.

In this paper we use an independent method to explore the viability of resonances and disc gap edges as sites for second generation planetesimal formation. We conducted three dimensional global disc models, using smoothed particle hydrodynamics (SPH), that include the two-fluid method of \cite{LaiPri2011a,LaiPri2011b}. In Section~\ref{sec:setup} we describe our computational method before presenting our results in Section~\ref{sec:results}, and discussing their implications in Section~\ref{sec:discussion}.

\section{Computational Method}
\label{sec:setup}

The calculations described herein have been performed using a three-dimensional SPH code. This SPH code has its origins in a version first developed by \citeauthor{Ben1990} (\citeyear{Ben1990}; \citealt{BenCamPreBow1990}) but it has undergone substantial modification in subsequent years. Energy and entropy are conserved to timestepping accuracy by use of the variable smoothing length formalism of \cite{SprHer2002} and \cite{Mon2002} with our specific implementation being described in \cite{PriMon2007}. Gravitational forces are calculated and neighbouring particles are found using a binary tree. Integration of the SPH equations is achieved using a second-order Runge-Kutta-Fehlberg integrator with particles having individual timesteps \citep{Bat1995}. The code has been parallelised by M. Bate using OpenMP and MPI.

Gas within the disc is subject to an artificial viscosity, implemented in a parameterised form as developed for SPH by \cite{MonGin1983}, which conserves linear and angular momentum, and was modified to deal with high Mach number shocks by \cite{Mon1997a}. We use the switch developed by \cite{MorMon1997} to reduce the action of artificial viscosity where the cause is not a shock. \cite{Pri2011} has shown that the use of such a switch is important in producing the high Reynolds numbers expected in astrophysical situations, and thus yielding realistic turbulence in subsonic and low Mach number flows.

\subsection{Disc model}

We model a protoplanetary disc with radial bounds for the gas component of $0.1 - 3$ \rorbit \ where \rorbit \ is the fixed orbital radius of the embedded protoplanet, taking a value of 5.2 au. The gas is modelled using $2 \times 10^6$ particles, as in \cite{AylBat2010}, but here we are not using radiative transfer. In addition to the gas component we model solid bodies ranging in radius from 10cm to 10m, straddling the metre-scale peak in gas drag \citep{Wei1977}, to examine their evolution. These bodies span a region from $0.4 - 2$ \rorbit \ ensuring that they are always immersed in the gas disc, and start with eccentricities and inclinations of zero. The solid bodies are modelled by $10^5$ solids particles.

 The equation of state used is that of a locally-isothermal ideal gas, where the disc maintains a temperature profile of the form $T \approx T_{\rm p}(r_{\rm p}/r)$, where $T_{\rm p}$ is the initial temperature at \rorbit \ ($T_{\rm p} \approx 73$K). The gas disc is established with a surface density profile of $\Sigma = \Sigma_{\rm p}(r_{\rm p}/r)^{1/2}$, where $\Sigma_{\rm p}$ is the gas surface density at \rorbit \ and takes a value of either $75$ (fiducial, or low density disc) or $750$ \sdensity \ (high density disc). The temperature and density profiles are chosen to match those of our previous work looking at protoplanetary discs \citep{AylBat2009, AylBat2010, AylBat2011}, where these values were taken from work that preceded us (\citealt*{LubSeiArt1999}; \citealt{BatLubOgiMil2003}). The temperature profile we have chosen is also equivalent to that used by both \cite{Paa2007} and \cite{LyrJohKlaPis2009} with whose work we draw comparisons in this paper. The solids follow the same surface density fall off as the gas, and the solids-to-gas surface density ratio in the overlapping region is 1 per cent (i.e. 1 per cent of the gas mass distributed between 0.4 - 2 \rorbit).

At the centre of the disc is a fixed potential representing a 1~\solarmass \ star. The planet is modelled as a potential on a fixed circular orbit, enwrapped by a surface force that allows accreted gas to pileup into a bound envelope \citep[see][]{AylBat2009}. The surface radius is set to 10 per cent of the planet's Hill radius, sufficiently small to allow the formation of a circumplanetary disc for high-mass protoplanets \citep{AylBat2009a,MarLub2011}. We implement an inner boundary for the disc that prevents gas flow onto the star, which due to the lack of a self-limiting mechanism would otherwise artificially rapidly drain material from the disc. The outer edge of the disc is bordered by ghost particles which represent the disc beyond 3~\rorbit \, preventing the disc from shear spreading into a vacuum \citep[for more details see][]{AylBat2010}. The initial discs were evolved in the absence of a planet or solid material until any transience resulting from settling had dissipated, which required just over 4 orbits of the disc's outer edge. Establishing a smooth, settled disc is of particular importance in these models given the tendency of solids to drift towards pressure maxima within a disc \citep{Whi1972}. The calculations discussed here were predominantly conducted without the inclusion of self-gravity, though it was introduced in some models to test the propensity of solid body accumulations to collapse under their own gravity.

\begin{figure}
\centering
\includegraphics[width=\columnwidth]{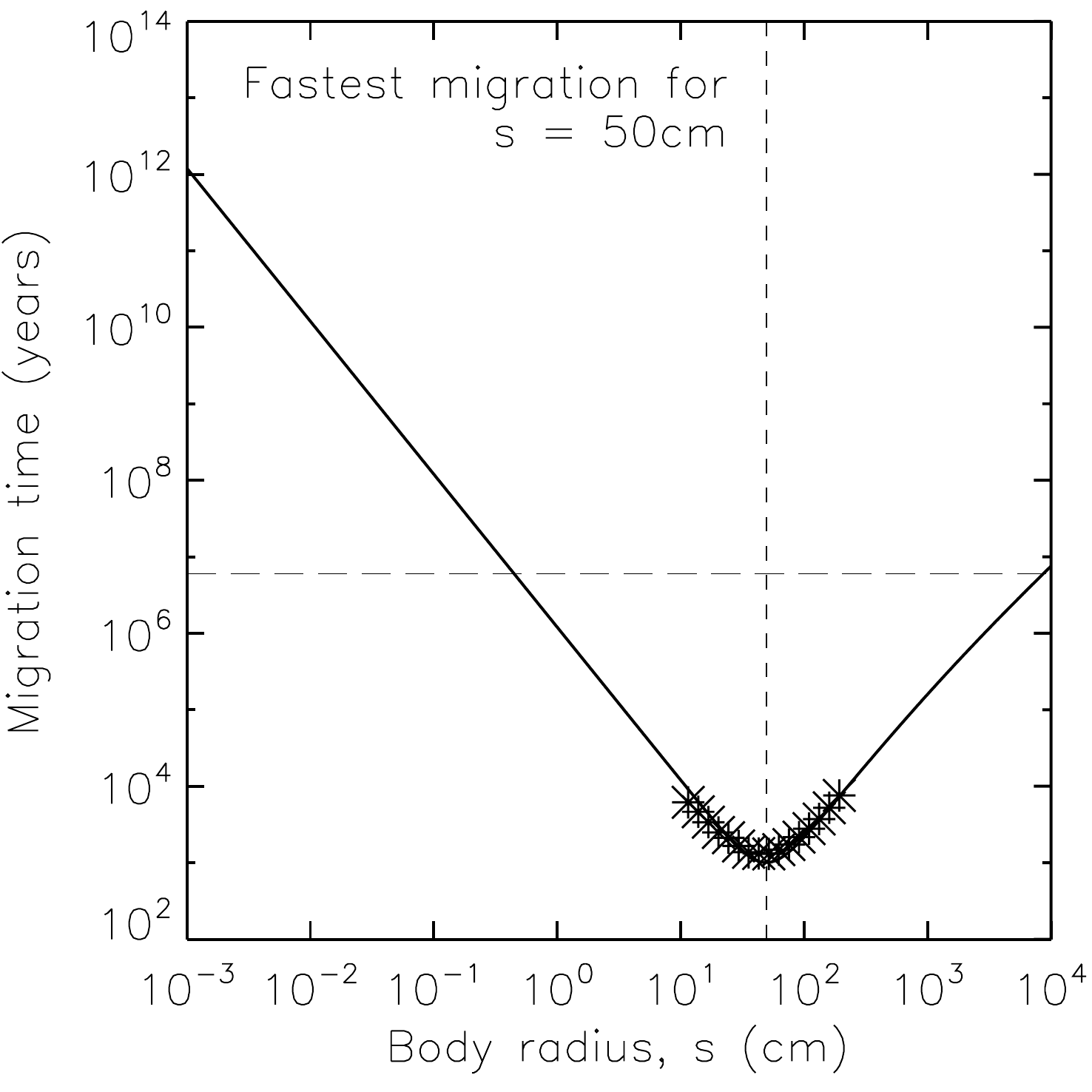}
\caption{Migration timescales for different sized bodies in an unperturbed disc (i.e. no protoplanet) at 5.2~au, where the gas surface density is 75~\sdensity. The solid line is obtained from analytic descriptions of gas drag \protect \citep{TakLin2002}, whilst the points marked are taken from our hydrodynamics calculations that span the size range for which drag-induced migration is most rapid. As can be seen, the numerical and analytic results are in good agreement. The horizontal dashed line marks a typical circumstellar disc lifetime \protect \citep*{HaiLadLad2001}, after which the absence of gas will end gas-drag induced migration of solid bodies. Only bodies of sizes ranging from $\sim$ 1cm to 100m can be expected to migrate significantly inwards during a disc's lifetime.}
\label{fig:timescales}
\end{figure}

\subsection{Gas-solids drag}
\label{sec:dragsheme}

Drag forces between the gas and solids are calculated using the explicit method developed for SPH by \cite{LaiPri2011a}. \citeauthor{LaiPri2011b} have derived an algorithm for simulating the evolution of gas and dust astrophysical mixtures using SPH that improves the accuracy of the previous algorithm developed by \cite{MonKoc1995}. The conservative part of the equation (pressure and buoyancy terms) is derived from a Lagrangian, whilst the form of the dissipative part (drag between the gas and dust) is as derived by \cite{MonKoc1995}. Densities for a given fluid are calculated by the usual SPH interpolation, using the standard bell shaped SPH kernel to sample neighbouring particles of the same type; this kernel is a cubic spline given as a function of $q$, where $q = r_{ij}/h_{ij}$, $r_{ij}$ is the separation of the particles and $h_{ij}$ is their average smoothing length. However, the drag interaction between the two fluids is calculated using a double-hump shaped kernel \citep{FulQui1996}; this is the standard SPH kernel multiplied by $q^2$. \cite{LaiPri2011a} found that computing the diagonal terms of the drag expression $(xx, yy, zz)$ could be achieved with small errors using the bell shaped kernel, but that the off-diagonal terms $(xy, xz, yz)$ were poorly normalised in this way. This situation was found to be enormously improved, by more than two-orders of magnitude, through use of a double humped kernel.

The scheme is implemented in a pairwise fashion. As SPH is a Lagrangian method, in which both gas and solids are modelled as free moving particles, the force calculated between the two is straightforwardly applied to both components, thus back reaction of the solids on the gas is inherently included and always present in our calculations. This formulation acts to conserve both linear and angular momentum, and energy, although our use of a locally-isothermal equation of state renders the energy conversation unimportant for our purposes. Tests exploring this numerical scheme are given in \cite{LaiPri2011a,LaiPri2011b}, with the latter considering the Stokes regime. In this paper the drag coefficient in the Stokes regime is taken from \cite*{BraDunLev2007}, whilst for the Epstein regime we use the form given by \cite*{BaiWilAse1965}. The transition from Stokes to Epstein occurs for $\lambda > 4s/9$ \citep{Wei1977,SteVal1996}, where $\lambda$ is the mean free path of a gas molecule and $s$ is the radius of the solid body (e.g. planetesimal or more likely dust in the Epstein regime).

In our test calculations the SPH particles representing the gas and solids populations were allowed independent smoothing lengths that were calculated for a given particle only by consideration of neighbouring particles of the same type. The smoothing lengths give SPH its natural resolution adaptability, with smaller smoothing lengths in regions of higher particle density. These tests revealed a process of artificial clumping for particles representing the solids, where this clumping resulted from insufficient resolution of the small scale structure in the gas field. This led to an essentially uniform drag force acting over solid particles assembled in regions of sub-gas resolution scale. Such congregating solids could assemble into extremely dense clumps due to the lack of a mutual pressure force to repel one another. A similar issue was found by \citet{PriFed2010} when using tracer particles in grid-based calculations of supersonic turbulence, and is also described for this SPH drag implementation in \cite{LaiPri2011a}, particularly Fig.~16. This clumping is obviously resolution dependent, as with sufficient gas resolution (precluding a perfectly smooth disc) there will always be structure to bring about slightly differing drag on neighbouring solids, and so avoid this mode of assembly. Two complementary steps were taken to remedy this problem. The first was to ensure high gas resolutions were used, that is to say a high ratio of gas-to-solids particles. This ensured that gas particles pervaded the distribution of solids particles, leading to differing distributions of gas particle neighbours for all the solid particles, such that their experienced drag was always somewhat different. The second measure was to calculate the gas-solids interactions using the gas smoothing lengths; the dust smoothing length is still defined, but plays no role in non-self-gravitating calculations.

\begin{figure*}
\centering
\includegraphics[width=0.98\textwidth]{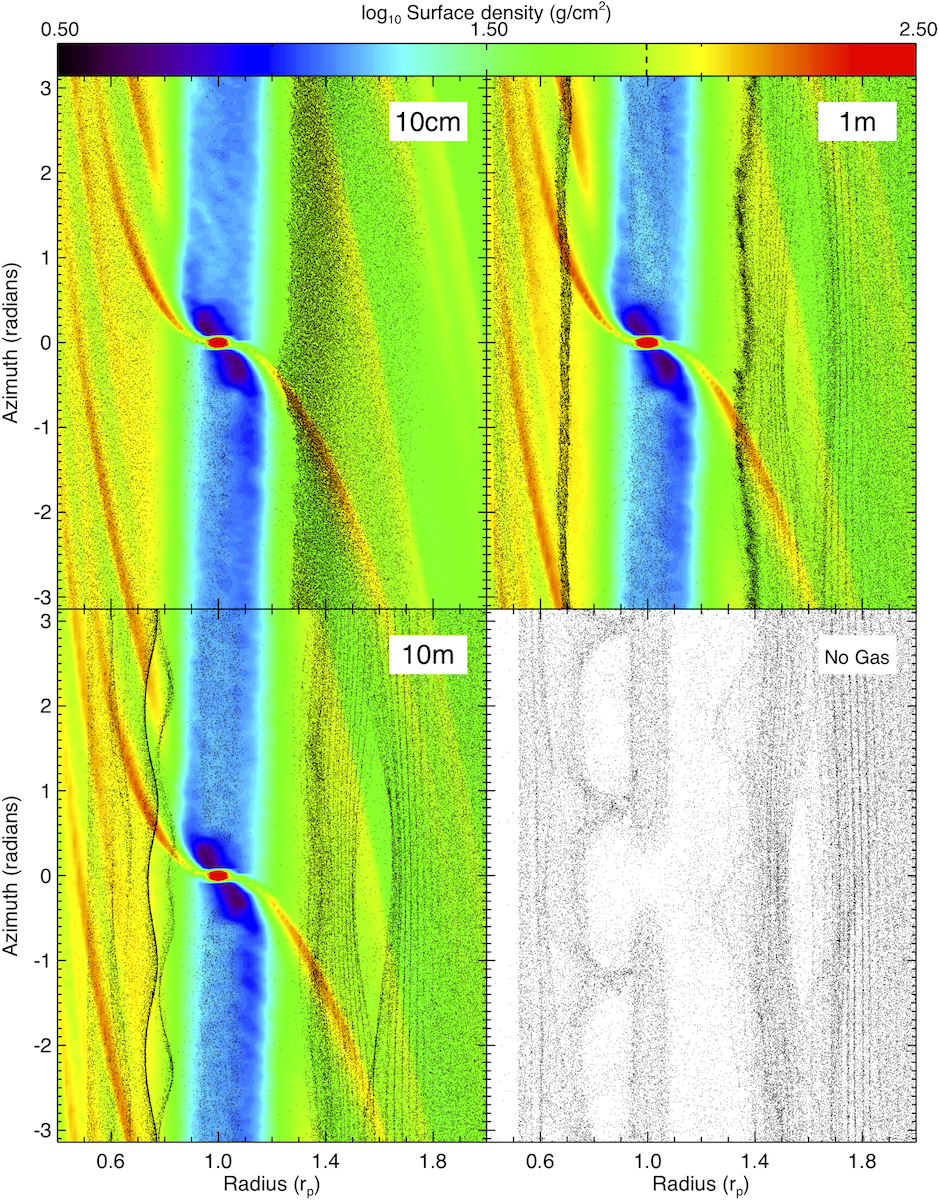}
\caption{Rendered images of gas surface density for a disc containing a planet of mass of 333~\earthmass, with an initial value of 75~\sdensity \ at \rorbit \ (marked in the colour scale by a dashed line). The calculations are performed in three-dimensions, employing cartesian coordinates, but here have been mapped to cylindrical polar coordinates. Solid bodies are plotted as points over the gas distribution. Solid bodies of radius 10cm, 1m, and 10m are considered (as marked). The lower right panel illustrates the solid body distribution that evolves in the absence of gas. The degree of structure (i.e. number of visible spiral ridges) seen in the solid body distribution increases with their size as a result of less significant gas drag. For 10cm grains, the solids much more closely follow the gas structure than the 10m bodies, which more closely resemble the no gas case.}
\label{fig:render75}
\end{figure*}

\begin{figure*}
\centering
\includegraphics[width=0.98\textwidth]{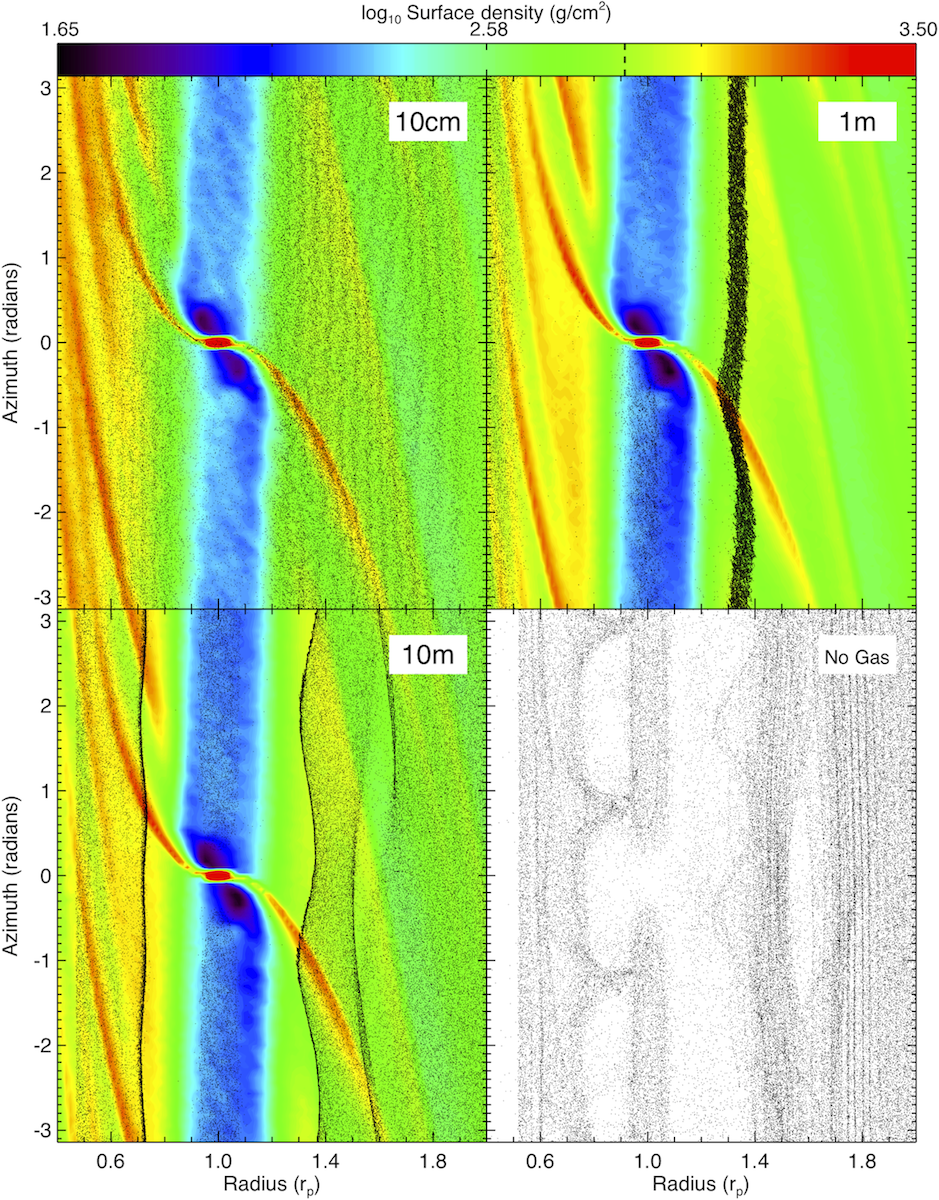}

\caption{As Fig.~\ref{fig:render75} but with an initial surface density of 750~\sdensity \ at \rorbit \ (marked in the colour scale by a dashed line). Due to the higher gas densities, the gas drag is considerably stronger in these cases, by a factor of $\sim 2$ for the 10m bodies in the Stokes regime, and by a factor of $\sim 10$ for the 10cm bodies, which are in the Epstein regime in the low density disc, and on the cusp between Stokes and Epstein in the high density disc. This may be equivalently stated as the stopping time of the solids is much shorter in the high density disc. As a result, the solid bodies that interact strongly with the disc (10cm - 1m) more rapidly respond to changes in the gas distribution. In particular, the 1m bodies quickly spiral in and accumulate in the pressure maxima at the outer edge of the gap in the gas disc opened by the protoplanet. Radial inspiral of 1m bodies due to drag allows the accumulation at the outer edge of the gap to grow rapidly.}
\label{fig:render750}
\end{figure*}

Use of this numerical scheme leads to excellent agreement with an analytical description of the drag-induced migration of solid bodies taken from \cite{TakLin2002}. This can be seen in Fig.~\ref{fig:timescales} in which radial migration rates measured from hydrodynamics models (and converted to migration times, $\tau = r_{\rm p}/\dot{r}$) for a whole range of solid body sizes are plotted over the analytic result given for our low density disc (we follow \citealt{Arm2007}'s application of \citeauthor{TakLin2002}'s method). The sizes modelled are such that they straddle the solid body radius for which migration is most rapid, which in this disc is a radius of 50cm.

\section{Results}
\label{sec:results}

\subsection{Planetesimal distribution}

Figs.~\ref{fig:render75} \& \ref{fig:render750} illustrate the distribution of solid bodies, of varying radii, on top of the gas surface density for discs with initial surface densities at \rorbit \ of 75~\sdensity \  and 750~\sdensity \ respectively. The lower right panel of both plots show the solid body distribution that develops in the absence of gas and gas drag. In all the cases shown, there is a 333~\earthmass \ planet embedded in the disc which has completed 50 orbits.

Without gas the solid bodies at large radii can be seen to develop a structure full of narrow ridges that result from their gravitational interaction with the planet. The density peaks that span the azimuthal range of these cylindrical polar plots are parts of a continuous peak that spirals away from the planet's location, most obvious at larger radii. An excellent fit to their distribution after 50 orbits can be made by a logarithmic spiral with a pitch angle of $\sim 2.5 \times 10^{-3}$~radians. This pattern is disrupted near orbital radii corresponding to planetary resonances for Keplerian orbits.

\begin{figure}
\centering
\includegraphics[width=\columnwidth]{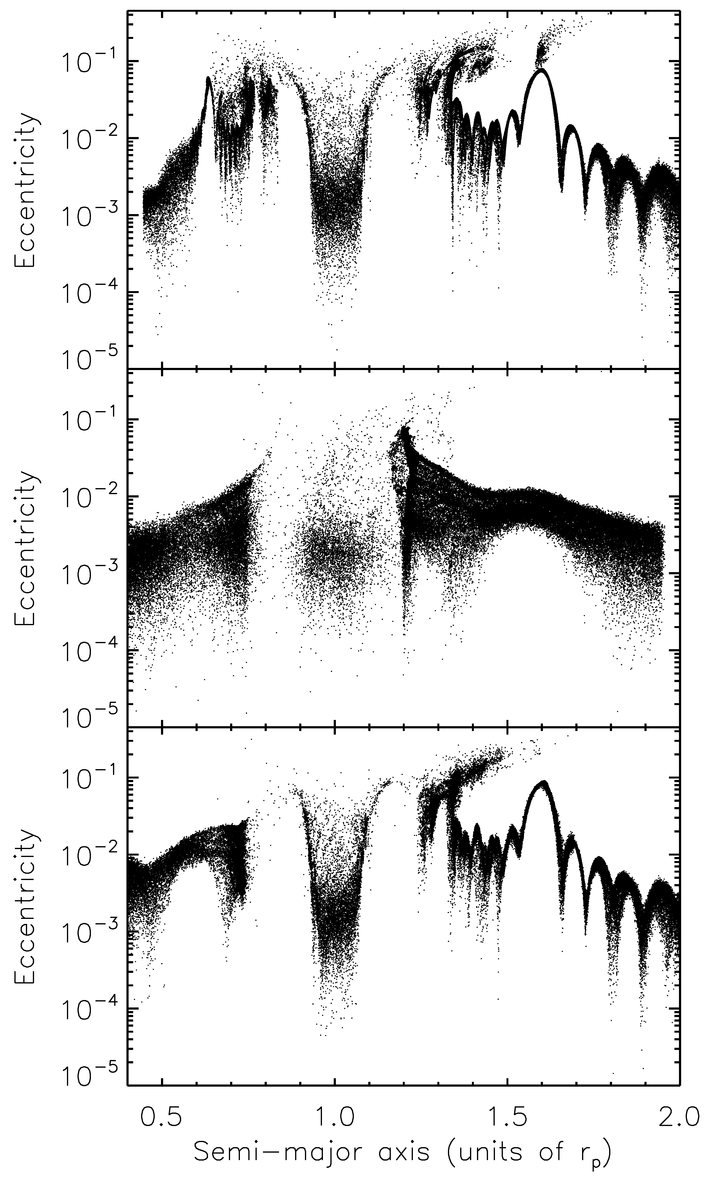}
\caption{Eccentricity of solids particles in an unperturbed low-density gas disc (top and middle panels); the solid particles are interacting with a Jupiter mass protoplanet {\em that the gas does not feel}. The top panel shows the distribution that develops for $s = 10$~m, whilst the middle panel is for $s = 1$~m in an identical disc. The drag induced radial drift of the 1~m bodies is sufficiently rapid that they are not captured in resonances with the planet, though there is a bulge in the eccentricity over the 2:1 resonance at $a \approx 1.6$~\rorbit. The 10~m bodies show a great deal of structure, in particular significant tufts over the 2:1 ($a \approx 1.6$) and 3:2 ($a \approx  1.3$) resonances that demonstrate trapping. The lower panel shows the distribution that develops for 1~m bodies {\em when the gas is also interacting with the protoplanet}. The density structure of the gas disc slows the drift of the 1~m bodies relative to the unperturbed case, allowing more significant excitation by the resonances.}
\label{fig:trapping}
\end{figure}

\cite{Paa2007} found that in a disc where the gas component was unperturbed by the protoplanet (i.e. gas doesn't feel the protoplanet's gravity, whilst the solids do), boulders with a relatively long stopping time could become trapped in planetary resonances. We replicated this scenario, and likewise found that our 10m bodies became trapped at, and suffered significant excitement of their eccentricities about the 2:1 and 3:2 resonances of a Jupiter mass planet, as can be seen in the top panel of Fig.~\ref{fig:trapping}. For our models involving 1m boulders, the gas drag induced radial velocities were sufficiently rapid to prevent capture in these resonances (middle panel Fig.~\ref{fig:trapping}), again in accord with \citeauthor{Paa2007}'s findings for bodies with short stopping times. The eccentricities of these rapidly infalling boulders are still stirred somewhat as they pass the resonances, however these eccentricities are then quickly damped by gas drag once the boulders move out of resonance. Following \cite{Paa2007}, we consider the work of \cite{WeiDav1985} who suggested the following equation for calculating the equilibrium eccentricity of a particle in a given mean motion resonance,

\begin{equation}
\bar{e} \approx \left(\frac{\Delta V/V_{kep}}{j + 1}\right)^{1/2},
\label{eq:ecc}
\end{equation}

\noindent where $j$ describes a resonance of the form $(j + 1):j$, and the $\Delta V/V_{kep}$ term describes the radial pressure support enjoyed by the gas that enables it to orbit with sub-Keplerian velocities. Using \citeauthor{WeiDav1985}'s estimated value of $\Delta V/V_{kep} = 5 \times 10^{-3}$ for the pressure support in a disc with a temperature profile of $T \propto r^{-1}$ gives $\bar{e} \approx 0.07/\sqrt{j + 1}$. This form suggests a mean eccentricity of $\approx 0.05$ at the 2:1 resonance, which compares favourably with the value for the 10~m bodies in an unperturbed disc measured over $a = 1.6 \pm 0.05$ of 0.057.

In our more usual calculations, where the planet's interaction is included with both the gas and planetesimal populations there is evidence of resonant trapping for the 1m and 10m solids bodies when we use our lower gas density (75~\sdensity \ at \rorbit), and for the 10m bodies only in the higher density (750~\sdensity) models, most noticeably in the 2:1 resonance. This is somewhat at odds with the results of \cite{FouMadGonMur2007} who state that accumulations outside the planet gap are unlikely to be associated with resonances, though they focus on the 3:2 resonance which is closer to the gap edge where the behaviour of solids is dominated by drag associated with the gas density structure. However, even at the 3:2 resonance we find eccentricity pumping that is similar to the unperturbed disc case, implying that both trapping at pressure maxima and resonances play a role. It of interest to note that the perturbation of the gas disc, and the planetesimal response to these perturbations (i.e. migration to pressure maxima) acts to slow the inward drift of 1m boulders compared with the case of the unperturbed gas disc. The result of this reduction in the net drift rate is to allow the eccentricity of 1m boulders to become much more strongly excited by the planet's resonances, which they had passed through quickly in the unperturbed gas disc; this can be seen by comparing the middle (gas unperturbed by protoplanet) and lower (conventional) panels of Fig.~\ref{fig:trapping}. The measured eccentricity in this 1~m case is 0.058, which again is in good agreement with the estimated mean given by equation~\ref{eq:ecc} for particles trapped in the 2:1 resonance. In all cases, the pumping of the eccentricities of these trapped bodies means that whilst they are concentrated in terms of semi-major axis, in real space they become somewhat under-dense as can be seen about $r \approx 1.6$~\rorbit \ in Fig.~\ref{fig:render75}. As such the capture at these resonant locations does not appear to be a promising site for the direct gravitational collapse of solids into larger bodies. The capture that we observe is qualitatively similar to that of \cite{Paa2007} who finds that larger bodies, those experiencing the least significant gas drag, are readily trapped in resonances, whilst the small bodies that are more well coupled to the gas are less prone to such capture.

The distribution of bodies in the no gas case (lower right panels of Figs.~\ref{fig:render75} \& \ref{fig:render750}, equivalent) at the planetary gap edge makes evident that highly eccentric orbits have developed both inside and outside the planet's orbit. The introduction of gas significantly damps the growth of these eccentricities. This is most evidently shown at the inner edge where a comparison between the no gas case with the neighbouring 10m bodies case (for which gas drag is weakest out of the shown scales) illustrates the immense difference in the resulting orbits for both disc densities. At large radii in the lower density gas disc the 10m and 1m bodies share a similar structure to one another and the no gas case, however at high gas densities this structure is smoothed out for 10m bodies, and the 1m bodies are entirely cleared from the region of interest due to rapid inward radial drift. The 10cm bodies are smoothed throughout the disc for gas discs of either density. Noticeably, the 10cm bodies in the lower density gas disc show evidence of drift and accumulation at the gap edge, whilst in the higher density disc, these small bodies are so well coupled to the gas that they adhere to the gas velocities, eliminating any effects of headwind and the associated inward drift.


\begin{figure}
\centering
\includegraphics[width=\columnwidth]{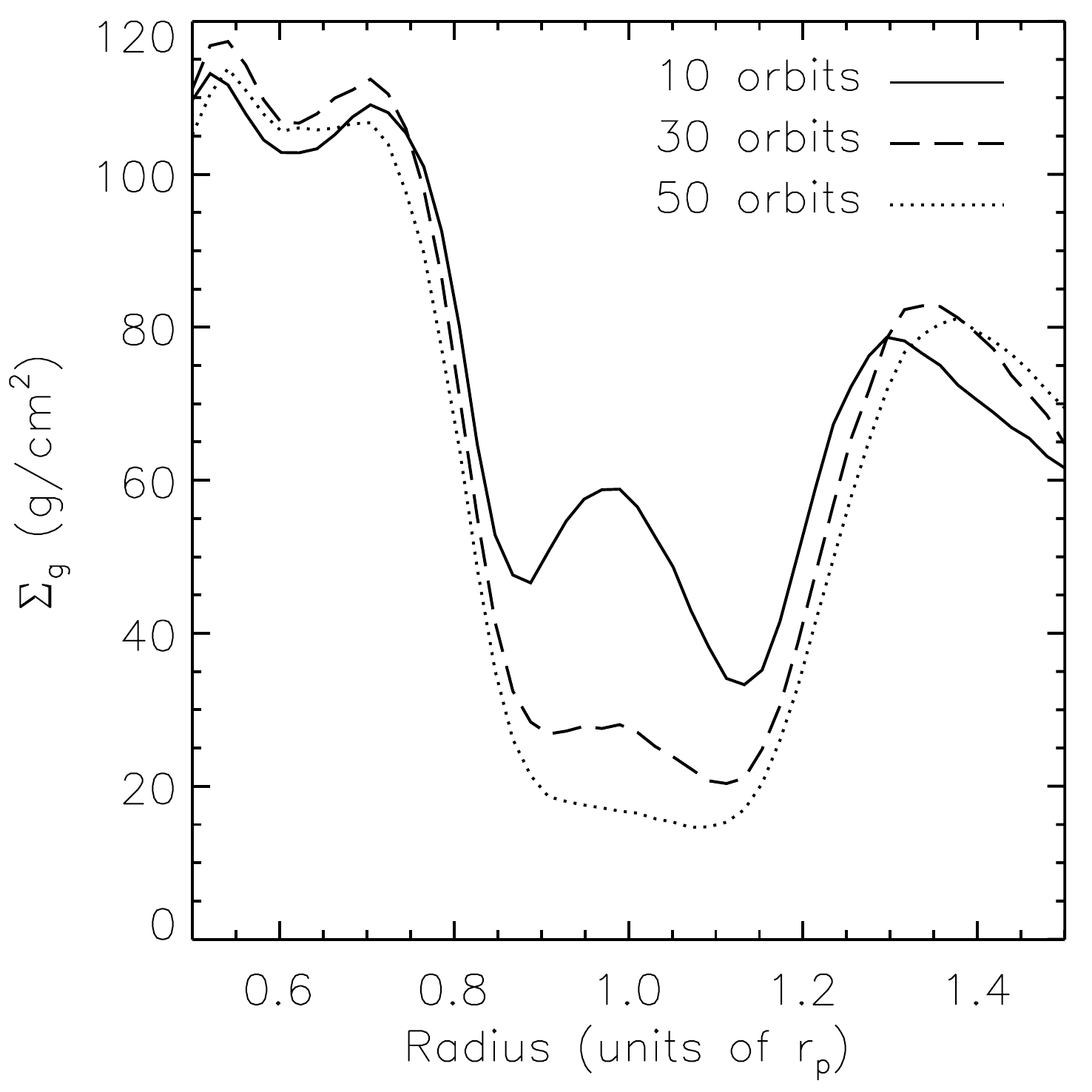}
\caption{Gas surface densities for a disc containing a Jupiter mass planet, where a region of azimuthal extent $\pi/2$ centred on the planet has been excluded to omit the protoplanet's contribution to the density in the corotation region. The surface density is shown for three periods as marked, illustrating the increasing evacuation of the gap with time, and the removal of the density peak that initially exists at \rorbit.}
\label{fig:gapopening}
\end{figure}

There is also a population of boulders trapped in the planet's corotation region (i.e. within the gas gap). These bodies migrate from regions approximately $r_{\rm p} \pm 0.1$ where the gas density drops off most rapidly as the gap opens, leaving a pressure maximum at \rorbit; this can be seen in Fig.~\ref{fig:gapopening}. Trapped particles will reside in the corbital region for many tens of orbits, however as Fig.~\ref{fig:gapopening} makes clear, after 50 orbits the gap is better evacuated, removing the pressure maxima from its centre (excluding the planet itself), but leaving a non-zero gas density. This residual gas will slowly act through drag upon the solid bodies, causing them to move towards smaller orbital radii and so eventually clear the gap, as was discussed by \cite{Paa2007}.

In the lower density case (Fig.~\ref{fig:render75}) the solid bodies of all scales can be seen to accumulate to varying extents at the gas density maxima marking the inner edge of the gap. However, for the high density gas disc (Fig.~\ref{fig:render750}) the 10cm grains show no signs of accumulating at any points in the disc excepting very slight over densities in the spiral arms associated with the planet. Instead these grains remain well coupled to the gas, not displaying significant radial drift over the course of 50 orbits. The 1m boulders, which do drift significantly, are unable to enter the gap etched by the Jupiter mass planet, leading to an enormous buildup of boulders at the gap's outer edge. In this case the gas density peak associated with the inner edge of the gap also slows the passage of inwardly drifting boulders, leading to a transient accumulation. However the gradient in the density at this peak ($r \approx 0.7$~\rorbit) which acts to drive the boulders outwards is relatively shallow, and thus is insufficient to fully arrest the inspiral of the boulders. As a result, the disc's inner region is slowly drained of material which cannot be replaced from larger radii due to the effective capture at the gaps outer edge.

\subsection{Solids pileup at the planetary gap}
\label{sec:pileup}

The torques exerted by a protoplanet embedded in a circumstellar disc that act to clear a gap in the gas, also act to clear solid bodies from the same region. In fact it has been found that gaps are more readily cleared in a dust population than in the gas \citep{PaaMel2004, PaaMel2006a, FouMadGonMur2007, MadFouGon2007}. Furthermore, \cite{PaaMel2004} found that a result of this clearing was a significant peak in the density of dust (1mm grains) just beyond the gap edge. This occurs as the solid material is forced out to the edge of the gap region. Subsequent models \citep{Paa2007, LyrJohKlaPis2009}, including those presented here have shown that this pileup of solids is prone to further evolution as it responds to the evolving density structure of a disc's gaseous component. For example, the opening of a gap in the gas disc leads to density maxima at the inner and outer edges, maxima to which solids are driven by gas drag. \cite{Paa2007} found that the outer density maxima was an effective barrier to the inwards drift of solid particles larger than $10 - 100 ~{\rm \mu m}$ (depending on the scale of the gas density peak, which is determined by the planet mass and disc properties). However, as was mentioned earlier (Fig.~\ref{fig:timescales}) only bodies ranging in radius from $\sim$ 1cm to 100m migrate significantly during the limited lifetime of the gas disc. In the discs considered here, the most significant migration occurs for solid bodies with radii in the decimetre to metre range. As they head towards the planetary gap these migrating bodies become held up by the gas density maxima at the outer edge, and to a much lesser extent by the gap clearing torques of the planet, which combine to prevent them from entering the corotation region.

Those solids that migrate the fastest accumulate most rapidly at the gap edge. We find that for 1m bodies the capture is $\sim 100$ per cent efficient. As a result, the rate at which mass collects at the gap edge is simply determined by the surface density of boulders, and their radial velocities due to gas drag. We can estimate this rate analytically and make comparisons with the pile up in the numerical simulations.

\begin{figure}
\centering
\includegraphics[width=\columnwidth]{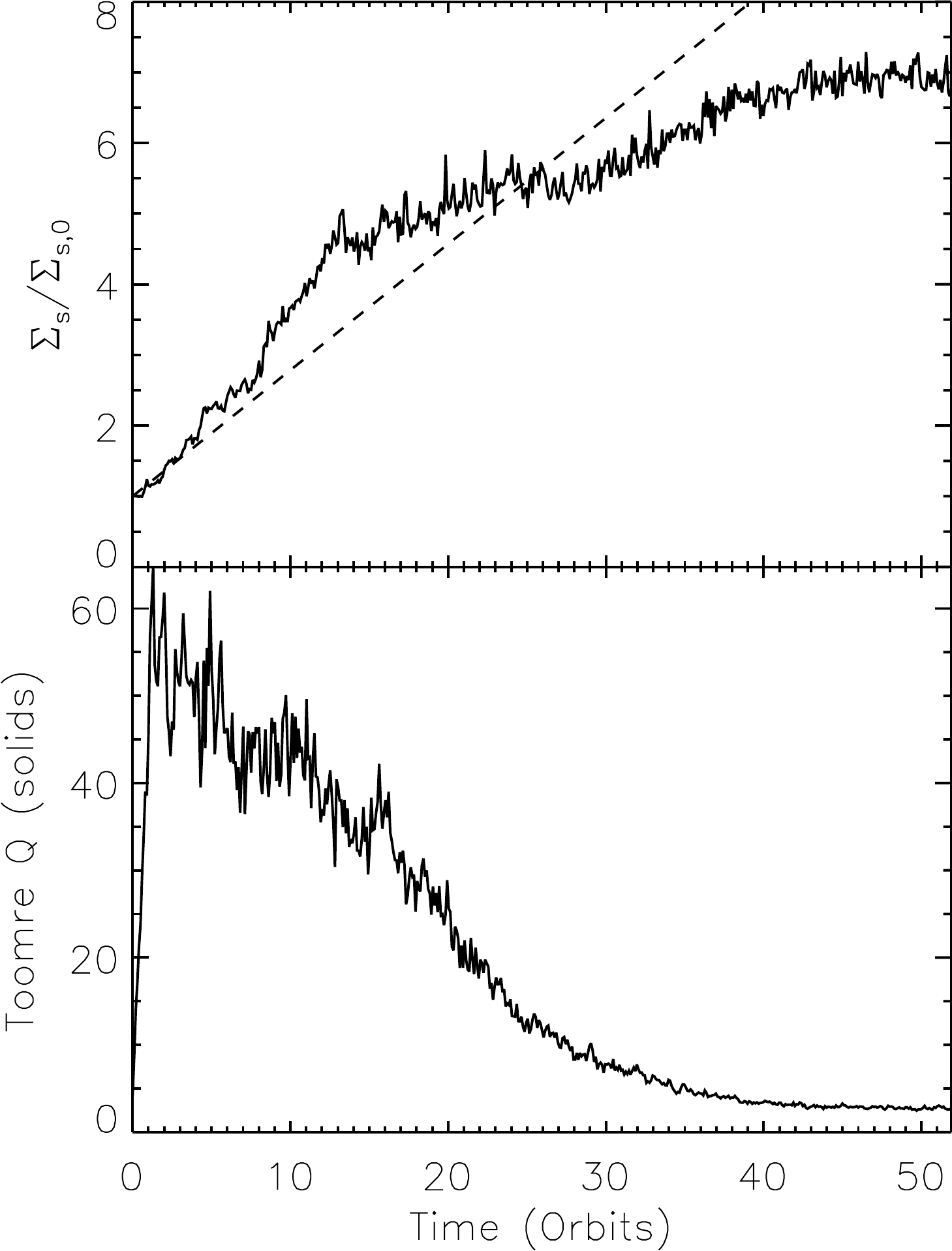}
\caption{For a disc of initial surface density 750~\sdensity \ at \rorbit, containing a Jupiter mass planet and solid bodies of 1m in radius. {\bf Top:} The rate of change of surface density of the solid body peak that forms at the outer edge of the planetary disc gap, given as a ratio to the initial unperturbed density of the same region ($1.25 - 1.35 r_{\rm p}$). The dashed line gives the analytic estimation from Section~\ref{sec:pileup}, whilst the solid line shows the measured change from the numerical model. {\bf Bottom:} The azimuthally averaged Toomre Q value for the solids at r = 1.3~\rorbit, the maximum of the growing solids density peak.}
\label{fig:rateofpileup}
\end{figure}

Defining some radius to denote the edge of the pileup region, here called ${\rm r_{peak}}$, we can calculate the flux of solids ($f_s$) crossing this boundary,

\begin{equation}
\frac{dM}{dt} = f_s = 2 \pi {\rm r_{peak}} \Sigma_{\rm s} ({\rm r_{peak}}) v_{rad},
\end{equation}

\noindent where $\Sigma_{\rm s}$ is the solids surface density, and $v_{rad}$ is the radial velocity of the solid bodies. To calculate the radial velocity we need an estimate for the drag acting on metre-sized bodies in the disc being considered; the following is based upon \cite{Arm2007} and \cite{Tak2009}. This drag is given by

\begin{equation}
F_d = 0.5 C_D \pi s^2 \rho_g v_{rel}^2,
\end{equation}

\noindent in which $s$ is the solid body radius, $\rho_g$ is the gas density, $v_{rel}$ is the relative velocity of the solid bodies with respect to the gas, and $C_D$ is the drag coefficient. This drag coefficient depends upon the Reynolds number, which in turn is dependent upon $\rho_g$, $v_{rel}$, $s$, and the sound speed $c_s$. The degree of drag can also be expressed by a stopping time, the calculation of which uses the mass of the solid body ($m_{\rm s}$), and is given by

\begin{equation}
t_{stop} = m_{\rm s} v_{rel}/ F_d.
\end{equation}

The relative velocity can be estimated assuming that the solid bodies are on Keplerian orbits, whilst the gas exists in sub-Keplerian orbits due to pressure support. The velocity difference that results from this pressure support in our disc setup is given by

\begin{equation}
v_{rel}  = -0.5 \eta v_{kep}
\end{equation}

\noindent where $v_{kep}$ is the Keplerian velocity, and $\eta$ is given by

\begin{equation}
\eta = -2.5 \frac{c_s^2}{v_{kep}^2} = -2.5 \left(\frac{H}{r}\right)^2.
\end{equation}

\noindent Using $v_{rel}$ to calculate the stopping time, we now have all the components required to calculate the radial velocity of a boulder (as was done for Fig~\ref{fig:timescales}), which is given by

\begin{equation}
v_{rad} = \frac{-\eta v_{kep}}{\frac{v_{kep}}{r}t_{stop} + \frac{r}{v_{kep}}t_{stop}^{-1}}.
\end{equation}

\noindent The radial velocity gives us the mass flux ($f_s$) into the region of the peak. Defining a width for the peak ($r_{\rm min}$ to $r_{\rm max}$) we find that the mean surface density of the peak should change as

\begin{equation}
\frac{d \Sigma_{\rm s}}{dt} = \frac{f_s}{A}
\end{equation}

\noindent where $A = \pi (r_{\rm max}^2 - r_{\rm min}^2)$. We apply this method to a high density disc, with an initial surface density at \rorbit \ of 750~\sdensity, $s = 1$m, and a peak width (determined from the hydrodynamical models) of 1.25 to 1.35~\rorbit. The dashed line in the top panel of Fig.~\ref{fig:rateofpileup} illustrates how this analytic method suggests solids surface density of the peak region should change with time. An important note is that this analytic description is only valid for a short period of evolution. Thereafter the solids arriving at the peak have travelled from substantially different orbital radii, which due to the disc's initial $\Sigma_{\rm s} \propto r^{-1/2}$ profile, provide less mass for a given $\Delta r$. Moreover, the midplane gas density goes as $\rho_{\rm g} \propto r^{-3/2}$, so that these solids migrate more slowly at large $r$ due to lower gas drag, and so arrive more slowly than this analysis suggests. Minding these limitations, the top panel of Fig.~\ref{fig:rateofpileup} also includes the measured change in the regions solids surface density, which shows the expected agreement with the predicted rate of change over the first $5-  10$ orbits.

This accumulation at the gap edge is of most interest if it continues to a point at which a gravitational instability can develop, causing a direct collapse of the solids to form larger bodies, be them planetesimals or fully fledged planetary cores. This was previously explored by \cite{LyrJohKlaPis2009} who found that Rossby wave instabilities (RWI) formed in the solids population (for certain size bodies) at the gaps outer edge, further concentrating the material, leading to collapse and the formation of approximately terrestrial mass planets. We see no evidence of circulation counter to the orbital flow in this region suggesting that the RWI is not at work in our models. We must therefore look to see if there might be a collapse purely as a result of the solids pileup due to inward drift. The bottom panel of Fig.~\ref{fig:rateofpileup} shows the evolution of the \emph{azimuthally averaged} Toomre Q value in the middle of the region of solids accumulation ($\approx 1.3$~\rorbit). This value is calculated using the formula of \cite{Too1964},

\begin{equation}
Q = \frac{\sigma_{\rm vs} \kappa_{\rm ep}}{\pi \Sigma_{\rm s} G},
\end{equation}

\noindent where $\sigma_{\rm vs}$ is the velocity dispersion of the solid bodies (their standard deviation), $\Sigma_{\rm s}$ is the solids surface mass density, $G$ is the gravitational constant, and $\kappa_{\rm ep}$ is the epicyclic frequency, which in a Keplerian disc is approximately equal to the angular velocity, $\Omega$. The early Toomre Q values are $>> 10$, but after 40 orbits this value has reduced to less than 3. A value of less than 1 would indicate the region had become prone to gravitational instabilities. Considering azimuthal variations in the structure of the pileup, around half ($\pi$) of the ring of 1m boulders in the high density gas disc appears likely to collapse, with a mean Toomre Q value of 0.3. The material comprising these unstable regions only resides therein for a little over 1 orbit due to the faster orbital velocity of the boulders compared to the unstable feature. Due to this transience it is not obvious from this analysis whether these regions will be able to collapse, however tests in which self-gravity was introduced led to their immediate collapse. It is difficult to say much more about these models as the collapse due to the introduced self-gravity prevents further evolution, but it can be noted that in cases where the self-gravity is present from the very beginning of the model the collapse occurs at earlier times. This suggests that such a vast reservoir of small boulders would not collapse in one event, but rather that as material arrived at the gap edge there would be numerous smaller collapses to build somewhat larger bodies.

An alternative means of checking the stability is to consider the Roche criterion. This effectively compares the gravitational shear due to the star acting on a collection of bodies with their own self-gravity. The Roche density is given by:

\begin{equation}
\rho_{\rm R} = \frac{9 M_{*}}{4 \pi a^3}
\end{equation}

\noindent where $M_{*}$ is the stellar mass, and $a$ is the semi-major axis at which the collection of bodies are in orbit \citep*{MicKokInu2010}. Collapse will occur if the local solids density ($\rho_{\rm s}$) is greater than the Roche density. Using this criterion the total mass of solids that exist in the 1m pileup in the high density gas disc that is prone to collapse is $\approx 5.3$~\earthmass \ after 50 orbits. Performing a clump search, where a clump is defined as a contiguous region that is more strongly self-bound than disrupted by gravitational shear due to the star, the largest such collection is found to have a mass of $\approx 3.4$~\earthmass. A somewhat smaller clump of 1.1~\earthmass \ is also found, as are several clumps with masses between $0.1 - 0.5$~\earthmass. We find no such significant bound clumps ($> 0.1$~\earthmass) in any other of our calculations. However, the 10cm bodies in the lower density gas disc have formed more than 10 clumps in excess of ${\rm M_{Lunar}/4}$, whilst the 1m bodies in a high density disc around a 100~\earthmass \ protoplanet (see Section \ref{sec:mass}) form hundreds of clumps with masses in excess of ${\rm M_{Lunar}}$. Noticeably, there are no clumps approaching such masses in the case of the 1m boulders in the 75~\sdensity \ gas disc.

\begin{figure}
\centering
\includegraphics[width=\columnwidth]{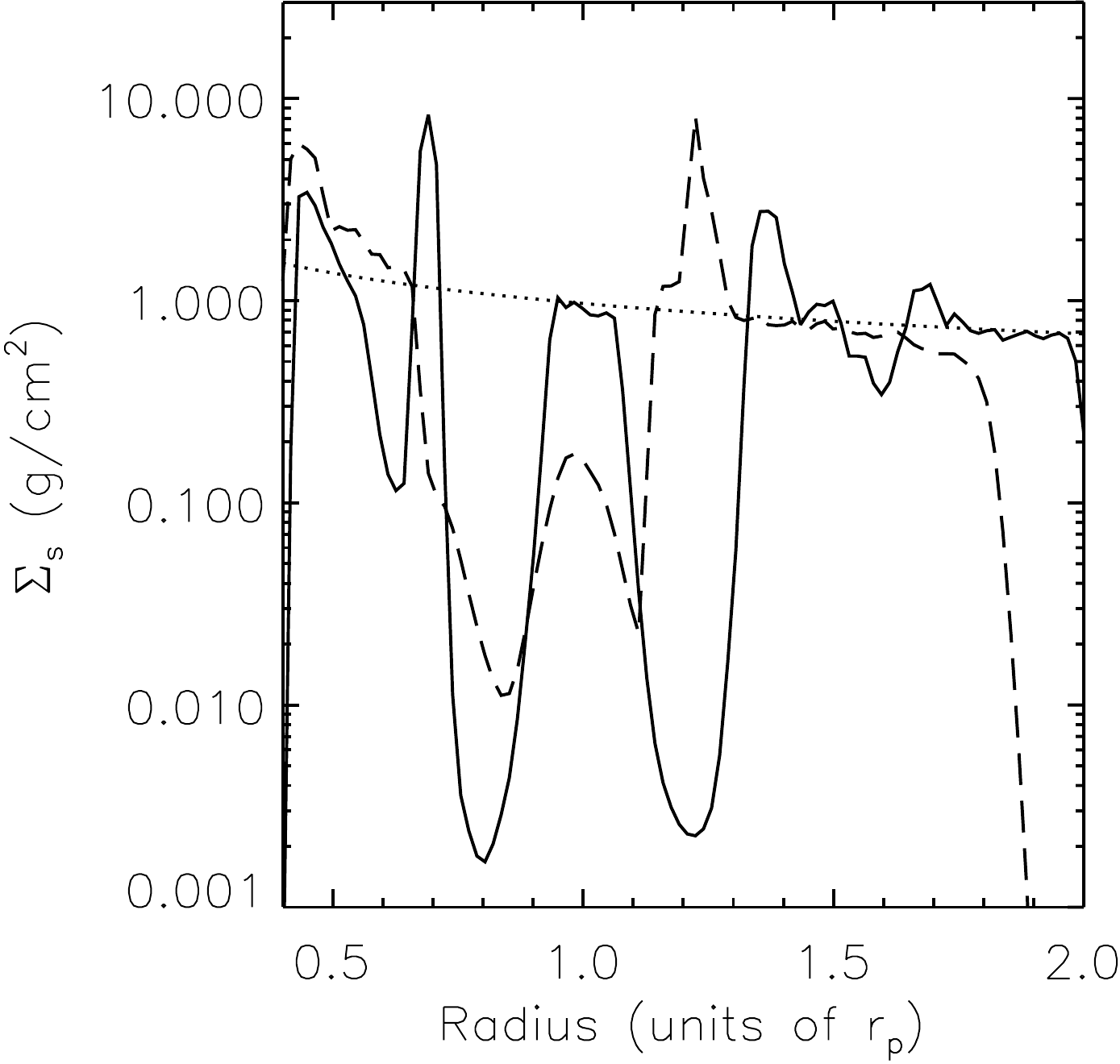}
\caption{Solids surface density profiles for a disc with an embedded 333~\earthmass \ protoplanet.  The solid line shows the profile that develops for the 1m solid bodies in our low density disc after 50 orbits. The dashed line is for the same solid body population but in a gas disc that that is not interacting with the protoplanet, such that the gas density remains smooth. The torques due to the protoplanet are sufficient to open a significant gap in the solid body population when the gas disc is smooth. However, the outer edge of this solids gap is closer to the protoplanet in the smooth disc as they are not driven out to a density peak in the gas at the gap edge. As such the solids gap in the unperturbed gas disc is an order of magnitude shallower in the region $r > r_{\rm p}$. The dotted line marks the initial unperturbed distribution of solid material.}
\label{fig:gapclear}
\end{figure}

We wanted to test the relative importance of the planetary torques acting to exclude 1m boulders and the gas drag induced migration to gas density maxima. The models which we had earlier performed to examine resonance capture, those in which the gas did not feel the influence of the protoplanet, give us an indication of how the migrating solid bodies behave without being driven to density maxima. The solids surface density that has developed after 50 orbits in such a disc is shown in Fig.~\ref{fig:gapclear} compared with a more usual perturbed disc. Clearly the protoplanet-solids gravitational interaction is sufficient to exclude significant material from the gap region, however the gap is narrower, and more sharply peaked at the outer edge. Moreover, the inner peak that appears in the conventional model is not present in unperturbed gas case as without a gas peak the material is not pinned at the inner edge, but rather drifts inwards unchecked.

\begin{figure*}
\centering
\includegraphics[width=0.98\textwidth]{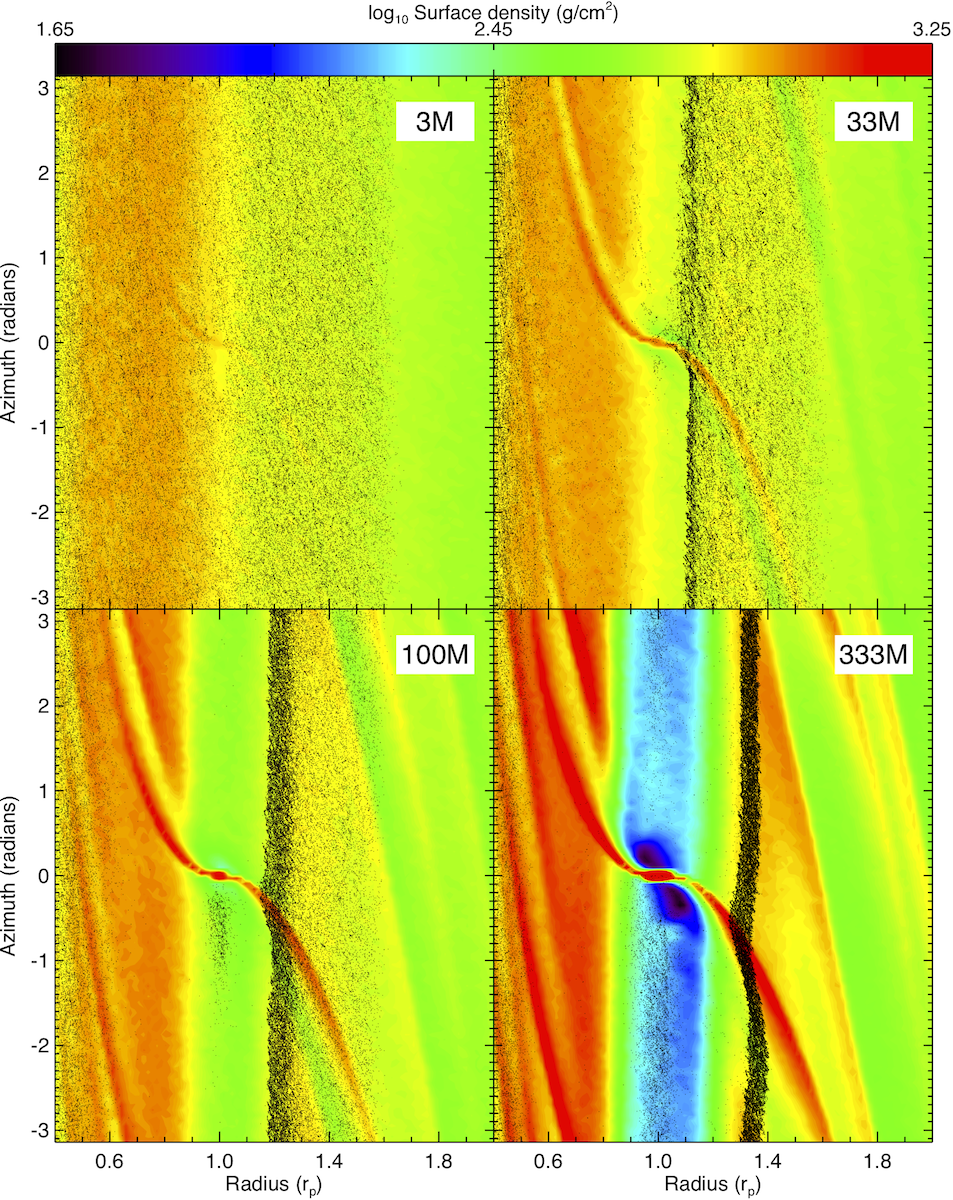}

\caption{Surface density maps and solid body distributions for discs with an initial surface density of 750~\sdensity \ at \rorbit, where the solid bodies are all 1m in radius. Protoplanets of mass 3, 33, 100, and 333~\earthmass \ (left to right, top to bottom as marked) are embedded in the discs which have been evolved for 50 planetary orbits. The pileup of the metre sized solid bodies can be seen clearly for the two highest mass planets. For the 333~\earthmass \ ($\sim$ Jupiter) case the pileup of solids occurs at the largest orbital radius of the four cases, and the draw of the pressure maxima has been sufficient to clear solids entirely outside of 1.4~\rorbit \ despite the rate of drag induced migration seen in the other cases leaving a population outside of this radius. For the 33~\earthmass \ case there is a slowing of the migration at the gap edge at 1.1~\rorbit, with some very slight leakage of boulders that may accrete onto the protoplanet or repopulate the inner disc ($0.6 - 0.9$ \rorbit).}
\label{fig:allmasses}
\end{figure*}

\subsection{Effect of planet mass}
\label{sec:mass}

\begin{figure}
\centering
\includegraphics[width=\columnwidth]{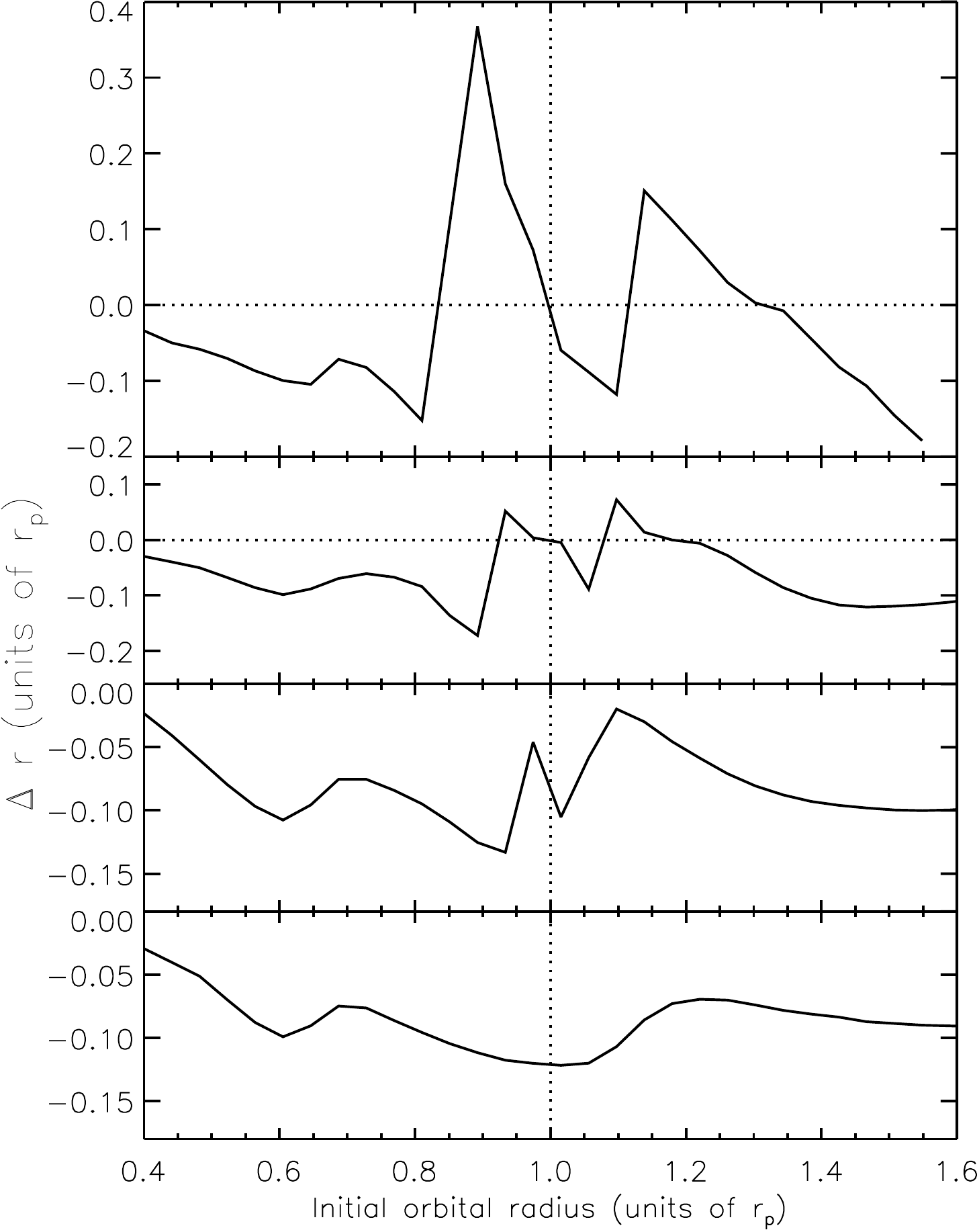}
\caption{Panels illustrating the average change in orbital radius of 1m solid bodies over the course of the last 10 of 50 orbits in high density discs containing planets of 333~\earthmass \ (top panel), 100~\earthmass \ (second panel), 33~\earthmass \ (third panel), and 3~\earthmass \ (bottom panel). The inward migration of these boulders can only be completely stopped by the two planets of highest mass; $\Delta r \geq 0$, above the horizontal dotted line. However, in the 33~\earthmass \ planet case the migration is still dramatically slowed and only $\sim 1$ per cent of the solid bodies drift inwards from the region $r > 1.0$ over the illustrated 10 orbits. The solids are held up at radii of $r \approx$ 1.3, 1.2, \& 1.1~\rorbit \ for the 333, 100, \& 33~\earthmass \ models respectively corresponding with the peaks in $\Delta r$ seen in the panels of this figure, and visible in Fig.~\ref{fig:allmasses}. The disc gaps opened in all but the lowest mass case possess sufficient density/pressure maxima at the edges, toward which the boulders are driven, to slow or prevent them from entering the gap. This allows the material to accumulate over long periods.}
\label{fig:rchange}
\end{figure}

We performed calculations with planets of four different masses to determine which of these could bring about the accumulation of metre-scale bodies in a way that might trigger planetesimal formation. The distribution of these boulders after 50 orbits in each case can be seen in Fig.~\ref{fig:allmasses}. The essential condition is that the migration of these boulders can be stopped at the gap edge, allowing accumulation to occur over a long period of time. The panels of Fig.~\ref{fig:rchange} show how the boulders move from their initial positions when evolving in a high density disc in the presence of a 333, 100, 33, or 3~\earthmass  \ protoplanet. The evolution is shown over the course of the final 10 planetary orbits of each model, each of which has completed 50 orbits in total. It can be seen that only 100 and 333~\earthmass \ planets are capable of completely stopping the inward drift due to aerodynamic drag in a 750~\sdensity \ gas disc. This is illustrated by the region of average outward migration just inside the peak of solids accumulation, with $\Delta r = 0$ occurring at the peak at $r \approx 1.3$~\rorbit \ for a 333~\earthmass \ protoplanet, and $r \approx 1.2$~\rorbit \ for a 100~\earthmass \ protoplanet. The 33~\earthmass \ planet is able to very significantly hold up the drift of the boulders, if not entirely stop it. The third panel shows a $\Delta r$ value that approaches 0 at a radius corresponding to the accumulation that can be seen in Fig.~\ref{fig:allmasses}, an accumulation that leaks just $\sim 1$ per cent of the boulders over 10 orbits. In the lowest mass case, the 3~\earthmass \ planet is not able to significantly modify the rate of inward drift experienced by the boulders (Fig.~\ref{fig:rchange}), and there is no sign of accumulation to be seen in Fig.~\ref{fig:allmasses}.

The higher the planet mass the more significantly it perturbs the disc, forming broader and clearer gaps with larger density/pressure maxima at either edge. The migration of boulders to these maxima is what arrests their inward drift. The potency of this process is illustrated for the 333 and 100~\earthmass \ protoplanets by the retention of 100 per cent of the boulders beyond $r > 1.0$ over the course of the final 10 orbits in these models. As stated above the 33~\earthmass \ protoplanet model loses $\sim 1$ per cent of these boulders, whilst the 3~\earthmass \ planet loses more than 11 per cent over the same period.

In the case of the 333~\earthmass \ protoplanet (top panel of Fig.~\ref{fig:rchange}) it is possible to see that solids have drifted outwards from inside of the protoplanet's orbit to become trapped in the coorbital region. Note that there can be no long term growth in solids density in the cororbital region as the supply of material is limited by the efficient trapping at the outer edge of the gap, and the inward drift and evacuation of the inner edge.

\subsection{A note about the drag scheme}

\begin{figure}
\centering
\includegraphics[width=\columnwidth]{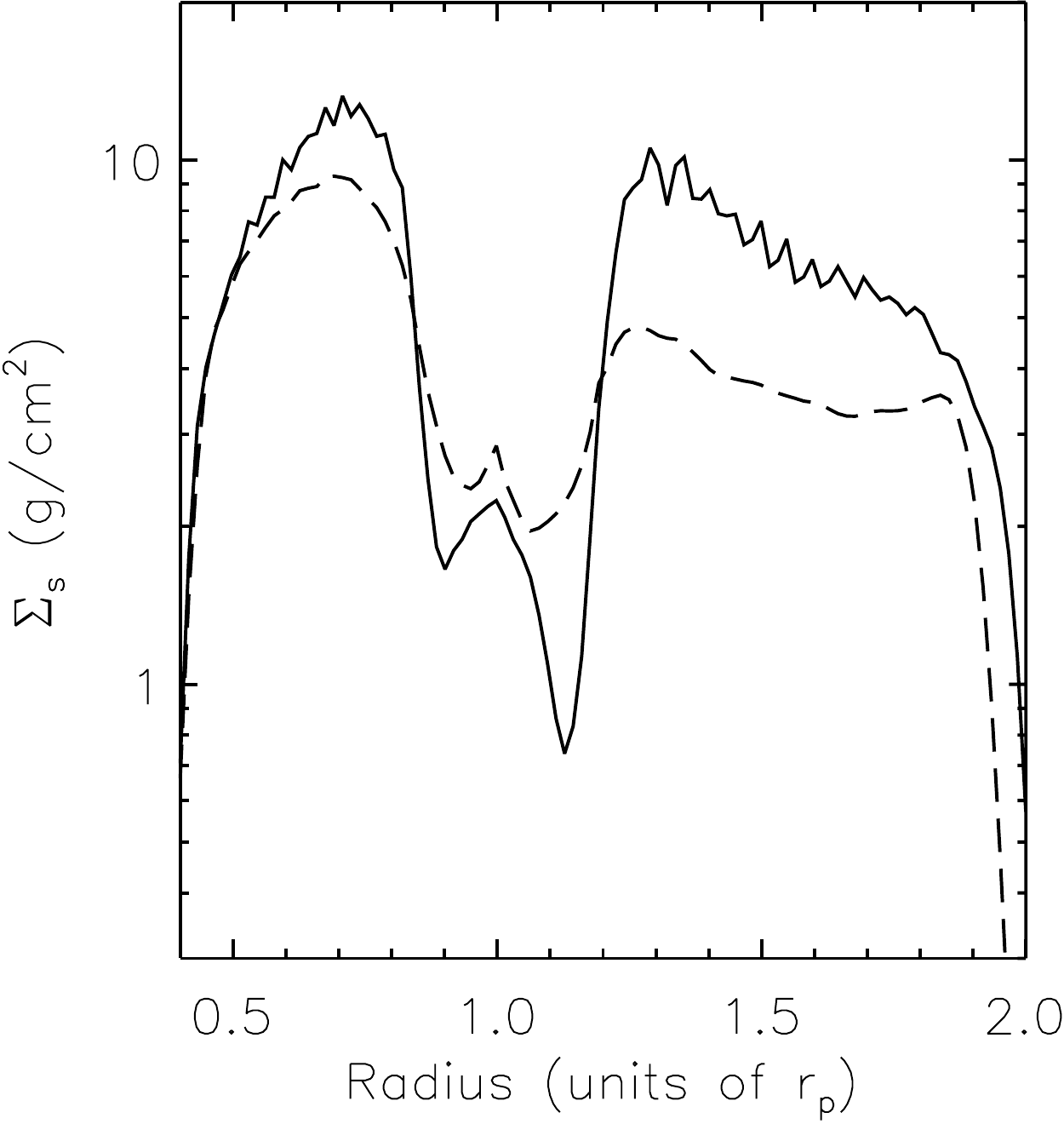}
\caption{Solids surface density in a high density disc about a Jupiter mass planet, where the solid material is constituted in 10~cm rocks. The solid line denotes the result of the typical calculations discussed here using \protect \cite{BraDunLev2007} and \protect \cite{BaiWilAse1965} to set the drag coefficient in the Stokes and Epstein regimes respectively. The dashed line gives the result for the same model where the drag coefficient is set using the continuous form given by \protect \cite{PerMur2011}.}
\label{fig:dragschemes}
\end{figure}

The drag calculations described in this work employ a drag coefficient, the value of which was set in the Stokes regime using \cite{BraDunLev2007} and in the Epstein regime using \cite{BaiWilAse1965} (as mentioned in Section~\ref{sec:dragsheme}). However, other schemes have been suggested, such as the continuous formulation given by \cite{PerMur2011} that stretches across both regimes. To test the impact of employing an alternative scheme, we performed a repeat of the $s=10$~cm solids model, embedded in a high density disc around a Jupiter mass planet. It is this combination of solids scale and gas density that leads to a significant degree of evolution on the cusp of the Epstein and Stokes drag regimes, and should therefore reveal differences due to the drag coefficient that arise in both regimes and across the transition. Fig.~\ref{fig:dragschemes} illustrates the surface density of solids that develops in both models, the solid line denoting our usual formulation, and the results of the \citeauthor{PerMur2011} scheme shown with a dashed line. The principle difference is that more material is able to flow across the gap using the latter scheme, which leads to an overall reduction in the mass of solids that persist in the disc, and so a reduction in the surface density across the radial range. This flow across the gap is enabled by higher values of the drag coefficient given by the \cite{PerMur2011} formulation, where these values differ particularly from our usual scheme in regions of lower gas density. As such the disc gap opened by the planet does not lead to such a significant drop in the drag coefficient, and functions less effectively as a barrier to radial drift. We must therefore note that the results found in calculations such as these can really only be compared with calculations employing the same, or very similar, drag schemes.

\section{Discussion}
\label{sec:discussion}

There have been a number of suggestions in recent years as to how the metre-size barrier to planetesimal growth might be overcome. Generally a process of trapping is required to increase the solids density, and so allow rapid formation by gravitational collapse to occur. This trapping may occur in vortices, or in gas density maxima to which solid bodies migrate. Previous work has shown that dust can be efficiently cleared from the vicinity of a protoplanet through gravitational interactions \citep{PaaMel2004, PaaMel2006a, FouMadGonMur2007, MadFouGon2007}. This exclusion from the corotation region, coupled with a continuous supply of additional material (i.e. due to inspiral of solids experiencing gas drag) can lead to a significant pileup of material, as discussed previously by \cite{Paa2007}, \cite{LyrJohKlaPis2009}, and \cite*{FouGonMad2010a}, with \citeauthor{LyrJohKlaPis2009} paying particular attention to the possibility of gravitational collapse within this accumulated solids population.

High mass planets (Saturn mass and above) open a clear gap in the gas disc, which results in gas density maxima at the inner and outer edges. Boulders are strongly driven towards these gas density maxima due to gas drag, as shown by \cite{Whi1972}, and as such accumulate in these regions. The combined effect of the torque interaction with the planet, that pushes the boulders away from the corotation region, and the action of gas drag is to almost entirely prevent boulders reaching the inner disc ($r < r_{\rm p}$). As such the accumulation of these solids can occur over long timescales, eventually reaching densities sufficient to trigger collapse, as we have seen about a Jupiter mass planet. \cite{LyrJohKlaPis2009} showed that Rossby wave instabilities in these regions could further concentrate the material, helping to trigger the formation of larger planetary emrbyos/planetesimals. However, we see no evidence of these instabilities in our models, most likely due to the relatively high viscosities involved, and the three-dimensional nature of the calculations that inhibits vortex survival \citep{LesPap2009}. Still, forgoing the contribution of RWI concentration, we still find clumps of material that are prone to collapse under their own self-gravity around the outer edge of the disc gap. \cite{LyrJohKlaPis2009} also found evidence of significant accumulation occurring in small regions about the L4 and L5 Lagrange points within the gap. Our models suggest that a majority of the material, after 50 orbits, is still well distributed along horseshoe orbits, not concentrating particularly about either of these Lagrange points. The only exceptions to this are in the high density gas disc, where the well coupled 10cm solids do not become trapped in horseshoe orbits at all, and the 1m bodies concentrate to an extent in forming a tadpole orbit around the L5 point (see Fig.~\ref{fig:render750}).

Given unlimited time it should be possible for large densities to accrue for many different sized solid bodies, but the effect is most rapidly seen for those that migrate inwards fastest, in the disc considered here this is around the metre scale ($s \approx 0.5$m, see Fig.~\ref{fig:timescales}). Recent numerical models by \cite{FouGonMad2010a} of circumstellar discs containing planets and a range of small grains, ranging in size from 100$\mu$m to 1cm, hint at this accumulation for such small bodies. Fig.~3 of their paper shows that the 1cm grains, which are the least coupled to the gas, are able to migrate inwards during their longest calculations, collecting to form a broad peak at the outer edge of the gap. Much smaller or larger bodies migrate inwards very slowly, such that only those in the range of $\sim$ 1cm to 100m will undergo any significant drift before the gas disc dissipates and the drag-induced migration ceases. Outside this range the solid bodies will exhibit no significant accumulation over the $\sim$ 6Myrs \citep{HaiLadLad2001} of the gas disc's existence. The range of sizes for solids that might appreciably drift and pileup could be increased if circumstellar disc lifetime's were found to be significantly longer, as has been suggested by \cite{Nay2009} whose observations and analysis suggest lifetimes of between $1.5 - 2$ times longer than those typically assumed.

It should be noted that the peak in solids that we find has a maximum density at $r \approx 1.3$~\rorbit \ for the 333~\earthmass \ ($\sim$ Jupiter mass) case, which is around the planet's 3:2 resonance. However, for a 100~\earthmass \ planet ($\sim$ Saturn mass), this peak is at 1.2~\rorbit, indicating that the position is a result of the gap width in the gas disc, which scales with mass, rather than a result of resonant interactions.

If a Jupiter mass planet does create planetesimals in its vicinity, it is possible that these bodies will undergo mutual interactions that throw them into close encounters with the planet itself. A result of such interactions could be to throw the planetesimals into the star or out to the very edges of the system, possibly providing a means of creating a distantly bound collection of planetesimals like the Oort cloud in our solar system. The larger mass clumps that form may collapse to create superearth mass terrestrial planets, or cores for a further generation of gas giant planets (though there ability to form would be severely constrained by the discs dispersal).

\section{Summary}

We have shown, in agreement with previous work, that a moderate to high mass planet, capable of opening a gap in its parent disc, will put the brakes on the inward radial migration of metre sized boulders. These bodies, migrating inwards from outside of such a planet due to gas drag, become trapped in the gas density maxima that forms at the edge of the disc gap. This capture is extremely effective, allowing the accumulation of these boulders to occur over long timescales, enabling solids densities to develop that may be prone to gravitational instabilities. As such, the outer edge of a planetary disc gap may function as a site of planetesimal or planet formation by gravitational collapse, overcoming the difficulty of collisionally growing beyond metre scales. Of course, this growth can only occur in the presence of an existing planet, which must have formed from a previous generation of planetesimals unable to grow in this way.

We found that there is a degree of resonant capture by planets of the inwardly migrating solids, in agreement with \cite{Paa2007}, but due to the eccentricity pumping of these resonantly trapped particles, they do not form dense clumps that might collapse. We have also shown that in a gas disc that is perturbed by the presence of a protoplanet, developing a non-smooth density structure, the drag acting on solid bodies can slow their drift in such a way as to promote their capture in planetary resonances.

\section*{Acknowledgments}

BAA and MRB would like to thank Monash University for hosting them as visitors from September 2010 $-$ July 2011. The calculations reported here were performed using the University of Exeter Supercomputer. Much of the analysis was conducted making use of SPLASH \citep{Pri2007}, a visualisation tool for SPH that is publicly available at http://users.monash.edu.au/~dprice/splash/. MRB is grateful for the support of a EURYI Award which also supported BAA. This work, conducted as part of the award `The formation of stars and planets: Radiation hydrodynamical and magnetohydrodynamical simulations'  made under the European Heads of Research Councils and European Science Foundation EURYI (European Young Investigator) Awards scheme, was supported by funds from the Participating Organisations of EURYI and the EC Sixth Framework Programme. GL and DJP are grateful to the Australian Research Council for funding via Discovery project grant DP1094585.

\bibliography{bibliography}

\end{document}